\documentclass{bmcart}
\usepackage{siunitx}
\usepackage{graphicx}
\usepackage{textcomp}
\usepackage{multirow}
\usepackage{cite} 
\usepackage{amsthm,amsmath}

\begin{document}
\begin{frontmatter}
\begin{fmbox}
\dochead{Research}

\title{Airborne Quantum Key Distribution with Boundary Layer Effects}

\author[
  addressref={aff1,aff2},                
  noteref={n1},                        
]{\inits{H.C.}\fnm{Hui-Cun} \snm{Yu}}
\author[
  addressref={aff3},
  noteref={n1},
]{\inits{B.Y.}\fnm{Bang-Ying} \snm{Tang}}
\author[
  addressref={aff4},
  ]{\inits{H.}\fnm{Huan} \snm{Chen}}
\author[
  addressref={aff1},
]{\inits{Y.}\fnm{Yang} \snm{Xue}}
\author[
  addressref={aff1,aff2},
]{\inits{J.}\fnm{Jie} \snm{Tang}}
\author[
  addressref={aff3},
]{\inits{W.R.}\fnm{Wan-Rong} \snm{Yu}}
\author[
  addressref={aff2},                   
  corref={aff2},                      
  email={liubo08@nudt.edu.cn}
]{\inits{B.}\fnm{Bo} \snm{Liu}}
\author[
  addressref={aff1},                   
  corref={aff1},                      
  email={slfly2012@163.com}   
]{\inits{L.}\fnm{Lei} \snm{Shi}}

\address[id=aff1]{
  \orgdiv{Information and Navigation College},             
  \orgname{Air Force Engineering University},         
  \city{Xi’an},                             
  \cny{China}                                    
}
\address[id=aff2]{
  \orgdiv{College of Advanced Interdisciplinary Studies},
  \orgname{National University of Defense Technology},
  \city{Changsha},
  \cny{China}
}
\address[id=aff3]{
  \orgdiv{College of Computer and Science},
  \orgname{National University of Defense Technology},
  \city{Changsha},
  \cny{China}
}
\address[id=aff4]{
  \orgdiv{College of Liberal Arts and Sciences},
  \orgname{National University of Defense Technology},
  \city{Changsha},
  \cny{China}
}

\begin{artnotes}
\note[id=n1]{These authors contributed equally to this work.}
\end{artnotes}
\end{fmbox}

\begin{abstractbox}

\begin{abstract} 
  With the substantial progress of terrestrial fiber-based quantum networks and satellite-based quantum nodes, airborne quantum key distribution (QKD) is now becoming a flexible bond between terrestrial fiber and satellite, which is an efficient solution to establish a mobile, on-demand, and real-time coverage quantum network. However, the random distributed boundary layer is always surrounded to the surface of the aircraft when the flight speed larger than \SI{0.3}{Ma}, which would introduce random wavefront aberration, jitter and extra intensity attenuation to the transmitted photons. In this article, we propose a performance evaluation scheme of airborne QKD with boundary layer effects. The analyzed results about the photon deflection angle and wavefront aberration effects, show that the aero-optical effects caused by the boundary layer can not be ignored, which would heavily decrease the final secure key rate.  In our proposed airborne QKD scenario, the boundary layer would introduce $\sim$\SI{3.5}{dB} loss to the transmitted photons and decrease $\sim$\SI{70.7}{\%} of the secure key rate. With tolerated quantum bit error rate set to \SI{10}{\%}, the suggested quantum communication azimuth angle between the aircraft and the ground station is within $60^\circ$. Furthermore, the optimal beacon laser module and adaptive optics module are suggested to be employed, to improve the performance of airborne QKD system. Our detailed airborne QKD  performance evaluation study can be performed to the future airborne quantum communication designs.
\end{abstract}

\begin{keyword}
  \kwd{Airborne Quantum Key Distribution}
  \kwd{Boundary Layer}
  \kwd{Aero-optical Effects}
  \end{keyword}

\end{abstractbox}
\end{frontmatter}

\section{Introduction}

Quantum key distribution (QKD), based on the fundamental principles of quantum mechanics, can provide information-theoretical-secure keys for distant users, with the capabilities of eavesdropping detection and tamper resistance~\cite{RN90,RN88,RN91,RN87,RN89}. Since the first BB84 protocol proposed~\cite{RN125}, QKD has shown broad and significant applications~\cite{RN104,RN67} in finance, government, and military. Currently, QKD systems both in fiber links~\cite{RN92,RN93,RN94,RN95} and free-space channels~\cite{RN97,RN96,RN103,RN99,RN98,RN101,RN100,RN102} have achieved substantial progress and have been gradually transferred from laboratory to realistic applications, such as the \SI{2000}{km} quantum communication backbone network between Shanghai and Beijing, an intercontinental quantum communication network among multiple locations on earth with a maximal separation of \SI{7600}{km}~\cite{RN65}, and an integrated space-to-ground quantum communication network over \SI{4600}{km}~\cite{RN66}, which shows that the quantum satellites can effectively expand the communication distance and construct ultra-long distance global quantum network. However, with constant orbits, limited communication time window and night-only quantum satellites, to construct a global-wide quantum-secure communication network is not an easy task. To establish a mobile, on-demand and real-time coverage quantum network, airborne QKD is an efficient solution~\cite{RN85,RN68,RN64,RN83,RN84,RN86,RN82}.

The first air-to-ground quantum communication demonstration was accomplished by Ludwig Maximilians University and the German Aerospace Center in 2013, with the platform flying at the speed of \SI{290}{km/h} and height of \SI{1.1}{km}. In 2020, Nanjing University reported an entanglement distribution based on drones which achieved \SI{200}{meters} coverage and duration of \SI{40}{minutes}~\cite{RN68}. Compared with satellite-to-ground quantum communication, airborne QKD features in high-speed maneuverability and suffers complicate atmosphere conditions, including atmospheric turbulence~\cite{RN110,RN107,RN109,RN106,RN108,RN105}, background noise~\cite{RN115,RN112,RN114,RN111,RN113} and attitude disturbance~\cite{RN116}. Furthermore, a very thin layer of air will stick over the surface of the aircraft with high velocity, resulting in the boundary layer (BL)~\cite{RN32}. The boundary layer would introduce random disturbance to the transmitted photons, which would reduce the coupling efficiency and fidelity of quantum states~\cite{RN33}. However, previous airborne QKD implementations only considered the influences from atmospheric turbulence and molecular scattering~\cite{RN82,RN86,RN33}, but ignored the boundary layer effects. S. Nauerth et al. concluded that the air swirl formed by the rotor wings would affect the transmission efficiency of communication channels in their air-to-ground quantum communication demonstration, but no further detailed analysis was presented~\cite{RN86}. When the aircraft is flying at a high speed, usually larger than \SI{0.3}{Ma}, the produced boundary layer will impair the performance of aircraft-based QKD~\cite{RN68}.

In this article, we propose a detailed performance evaluation scheme of airborne QKD with boundary layer effects. We firstly propose an air-to-ground QKD scenario with decoy BB84 protocol. Then, the photon deflection angle is evaluated by estimating the reflection index distribution of the surrounded boundary layer and performing the ray-tracing method. Afterwards, the Strehl Ratio caused by wavefront aberration of quantum signal states is evaluated by calculating the optical path length ($OPL$) and optical path difference ($OPD$). Finally, the overall photon transmission efficiency, quantum bit error rate and final secure key rate can be estimated. With common experimental settings, the boundary layer would introduce $\sim$\SI{3.5}{dB} loss to the transmitted photons and decrease $\sim$\SI{70.7}{\%} of the secure key rate, which shows that the aero-optical effects caused by the boundary layer can not be ignored. With tolerated quantum bit error rate set to \SI{10}{\%}, the suggested quantum communication azimuth angle between the aircraft and the ground station is within $60^\circ$. Furthermore, the beacon laser module and adaptive optics module are suggested to be employed, to improve the performance of airborne QKD system. Our detailed airborne QKD performance evaluation study can be performed to the future airborne quantum communication designs.

\section{Preliminaries}

\subsection{Decoy state quantum key distribution}

The most implemented protocol in realistic QKD systems is decoy state protocol, which can efficiently defense the photon number splitting attacks and can perform the weak coherent photon source to replace the single photon source in the implementations. The decoy state QKD protocol has been widely performed in the fiber-based, satellite-based and airborne-based QKD systems.

Thus, in this article, we introduce the vacuum and weak decoy BB84 protocol in the following QKD scheme with boundary layer effects~\cite{RN47}, where the final secure key rate can be calculated as 
\begin{equation}
\label{equ_R}
    R \geq q\left\{Q_1\left[1-H_2\left(e_1\right)\right] - Q_{\mu} f\left(E_{\mu}\right) H_2\left(E_{\mu}\right)\right\},
\end{equation}
where $Q_1$ is the gain of the received single photon states, $e_1$ is the error rate of single photon states, $f(x)$ is the information reconciliation efficiency for correcting error bits, $\mu$ is the intensity of the signal state. $Q_\mu$ and $E_\mu $ represent the gain of signal states and the overall quantum bit error rate (QBER) respectively. $H_2(x)$ is the binary Shannon entropy, which can be calculated as
\begin{equation}
    H_2(x) = -x\log(x) - (1-x)\log(1-x).
\end{equation}

Given the photon transmission efficiency $\eta$, $Q_\mu$ is calculated as
\begin{equation}
\label{equ_Qmu}
    Q_{\mu}=Y_{0}+1-e^{-\eta \mu},
\end{equation} 
where $Y_0$ is the dark count rate of QKD systems. Thus, the error gain of signal quantum states can be given by
\begin{equation}
    E_\mu Q_\mu = e_0Y_0 + e_d (1 - e^{-\eta\mu}),
\end{equation}
where $e_0$ is the error rate of dark counts, usually $e_0 = 0.50$. $e_d$ is the misalignment error rate of QKD systems.

Thus, the quantum bit error rate $E_\mu$ can be calculated as
\begin{equation}
\label{equ_Emu}
    E_\mu = {E_\mu Q_\mu}/Q_\mu .
\end{equation}

The gain of single photon states $Q_1$ can be  calculated as
\begin{equation}
\label{equ_Q1}
    Q_{1} \geq Q_{1}^{L, \nu, 0}=\frac{\mu^{2} e^{-\mu}}{\mu \nu-\nu^{2}}\left(Q_{\nu} e^{\nu}-Q_{\mu} e^{\mu} \frac{\nu^{2}}{\mu^{2}}-\frac{\mu^{2}-\nu^{2}}{\mu^{2}} Y_{0}\right),
\end{equation}
where $L$ denotes the lower bound value, $\nu$ is the intensity of decoy photons, $Q_\nu$ is the gain of decoy states. 

The error rate of single photon states $e_1$ can be calculated as
\begin{equation}
\label{equ_e1}
    e_{1} \leq e_{1}^{U, \nu, 0}=\frac{E_{\nu} Q_{\nu} e^{\nu}-e_{0} Y_{0}}{Y_{1}^{L, \nu, 0} \nu},
\end{equation} 
where $Y_{1}^{L, \nu, 0}$ is the yield of single photon states
\begin{equation}
    Y_{1}^{L, \nu, 0} = \frac{Q_1}{\mu e^{-\mu}}.
\end{equation}

The error gain of decoy states $E_\nu Q_\nu$ can be calculated as
\begin{equation}
    E_{\nu} Q_{\nu}=e_{0} Y_{0}+e_{d}\left(1-e^{-\eta \nu}\right).
\end{equation}

\subsection{Aero-optical effects}

In the airborne QKD procedure, aero-optical effects will be introduced to the photons, which are propagated through the density-varying flow field of the boundary layer. Typical aero-optical effects mainly include wavefront aberration, jitter, intensity attenuation and so on.

\begin{figure}[h!]
    \centering
    \includegraphics[width=0.8\textwidth]{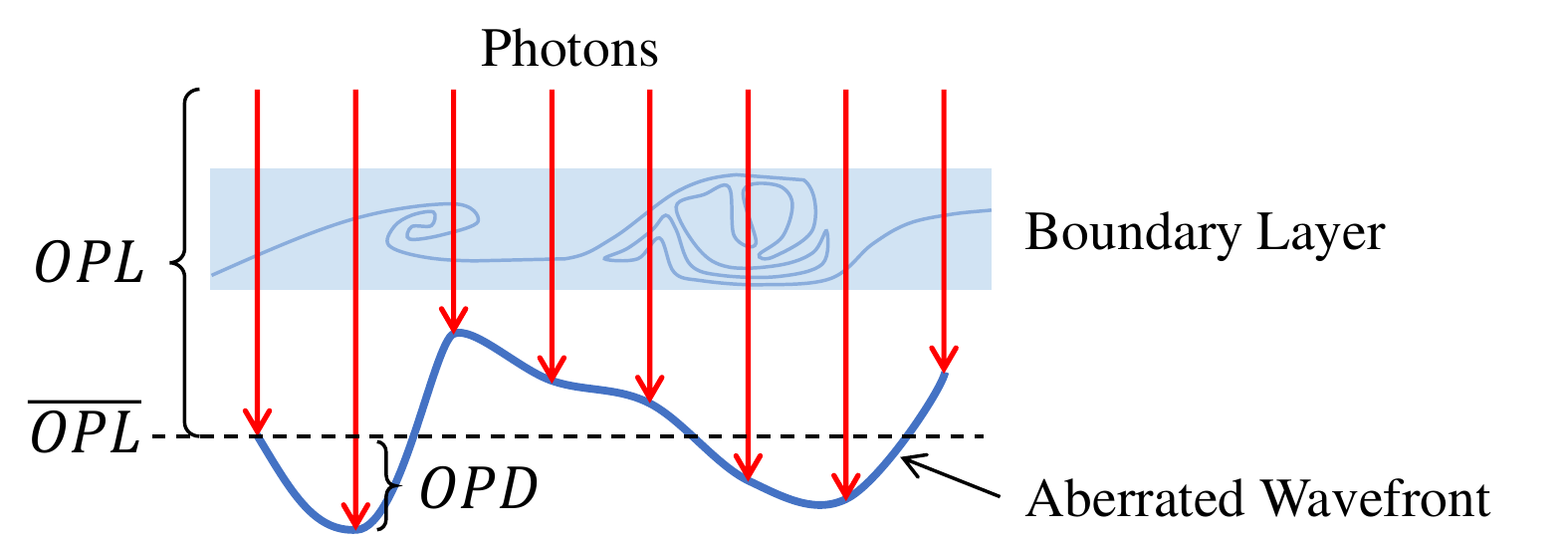}
    \caption{\label{fig1}Schematic diagram of the aberrated wavefront.}
  \end{figure}

Relevant parameters of aero-optical effects are optical path length (\textit{OPL}), optical path difference (\textit{OPD}) and Strehl Ratio (\textit{SR}), as shown in Figure~\ref{fig1}~\cite{RN2071}.

The aero-optical effects are fundamentally caused by the gradient refractive index $n$ due to the variable-density flow field, which is expressed by the Gladstone-Dale equation~\cite{RN74}
\begin{equation}
\label{equ_n}
    n = 1 + \rho K_{GD},
\end{equation}
where $\rho$ is the density of flow field. $K_{GD}$ is the Gladstone-Dale constant decided only the wavelength $\lambda$~(\textmu m) of photons~\cite{RN74}
\begin{equation}
K_{GD} = 2.23 \times 10^{-4} \times \left( 1 + \frac{7.52 \times 10^{-3}}{\lambda^2} \right).
\end{equation}

The refractive index field of the airborne boundary layer can be calculated by dividing the density field $\rho$ into sufficiently small squares and performing the Gladstone-Dale equation. The scattered photon path $P$ through the boundary layer can be calculated by performing the ray tracing methods~\cite{RN76,RN36}.

$OPL$ of the photons is calculated by integrating the refractive index $n$ along the propagation path $P$~\cite{RN122,RN123}.
\begin{equation}
OPL(x,y,t) = \int_{P} n(x,y,t) d p.
\end{equation}

$OPD$ shows the configuration of the wavefront and is defined as
\begin{equation}
OPD(x,y,t) = OPL(x,y,t) - \overline{O P L}.
\end{equation}

The overline denote the spatial average over the optical aperture. And the phase difference of photons can be defined by
\begin{equation}
\label{equ_phi}
\phi = \frac{2\pi OPD}{\lambda}.
\end{equation}

\section{Airborne QKD with boundary layer effects}

\subsection{airborne QKD Scenario}

\begin{figure}[h!]
  \includegraphics[width=0.8\textwidth]{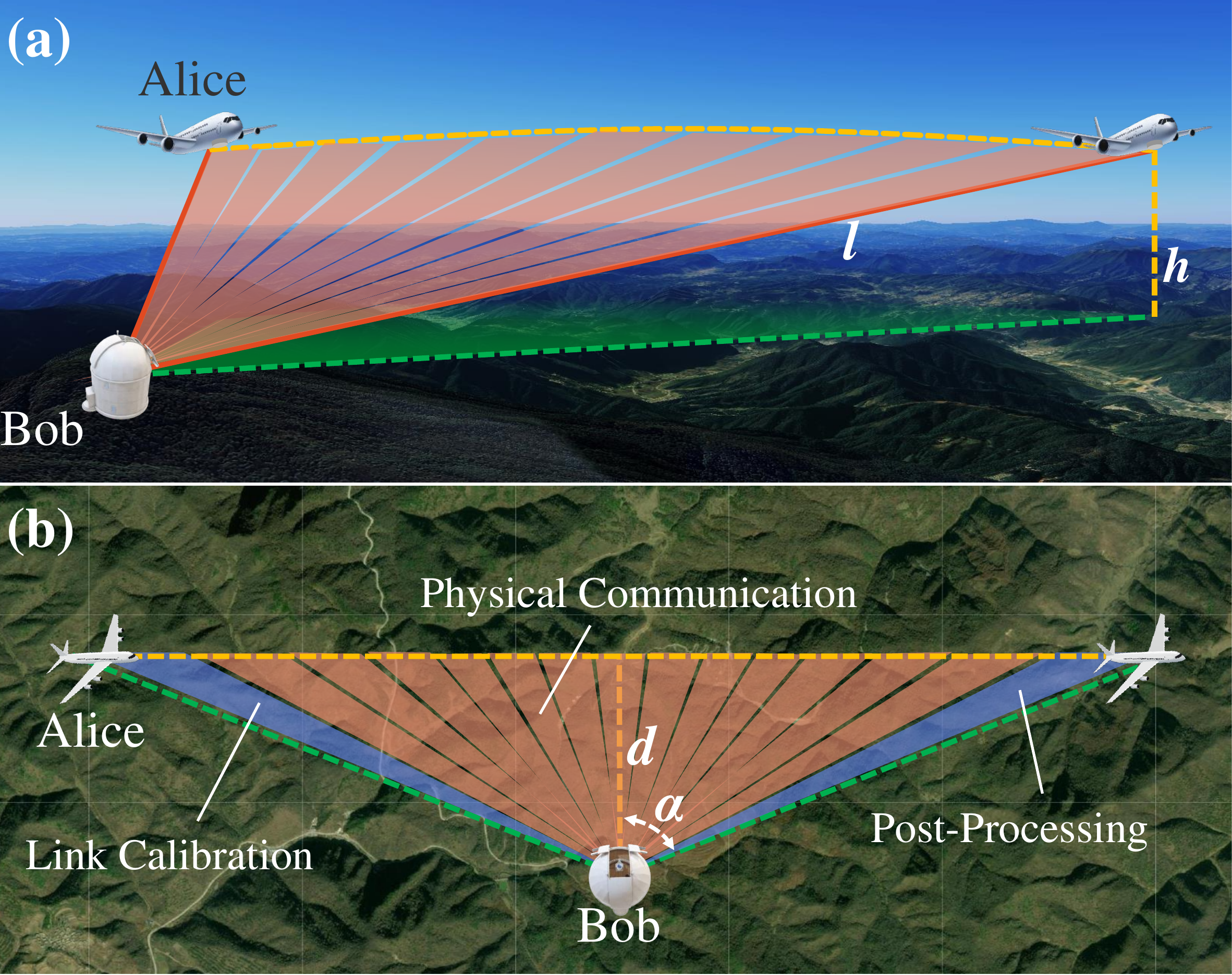}
  \caption{\label{fig2}(a) Main view and (b) top view of schematic diagram of downlink airborne QKD. The aircraft (Alice) flies in a certain path obliquely above the receiving ground station (Bob).}
\end{figure}

The air-to-ground QKD scenario is shown in Figure~\ref{fig2}, the quantum photon source is fixed in the airfoil of the aircraft (Alice) and the QKD receiving module is located at the optical ground station (Bob). Assume that Alice is flying with a constant velocity $\vec{v}$. The positions of Alice and Bob in East-North-Up (ENU) coordinate system are $S(x_s,y_s,z_s)$ and $O(x_0,y_0,z_0)$. Thus, the distance $l$ between Alice and Bob is
\begin{equation}
    l = \left| \overrightarrow{SO} \right|.
\end{equation}

The relative flying height $h$ can be calculated as
\begin{equation}
    h = \left| \overrightarrow{SO} \cdot \vec{Z}\right|,
\end{equation}
where $\vec{Z}=(0,0,1)$ is the unit vector of Z axis of ENU coordinate system.

In the top view of airborne QKD scheme (shown in Figure~\ref{fig2}(b)), the shortest horizon distance $d$ between Alice and Bob can be calculated as
\begin{equation}
    d = \frac{\left|(v_x,v_y,0)\cdot (SO_x,SO_y,0)\right|}{\left|(v_x,v_y,0)\right|}.
\end{equation}

The relative azimuth angle $\alpha$ between Alice and Bob can be calculated as
\begin{equation}
    \alpha = \arccos\left(\frac{(v_x,v_y,0)\cdot (SO_x,SO_y,0)}{\left|(v_x,v_y,0)\right|\left|(SO_x,SO_y,0)\right|}\right) - \frac{\pi}{2}.
\end{equation}

In the airborne QKD scheme, we perform weak-vacuum decoy BB84 protocol with signal photon intensity $\mu$ and decoy photon intensity $\nu$. The modulating probability of signal (decoy) states is $P_s$ and $P_d$. The airborne QKD scheme mainly includes three procedures: link calibration, physical communication and post-processing procedure. Once the quantum communication link is established between Alice and Bob after the link calibration procedure, modulated photons are transmitted from Alice to Bob during the physical communication procedure and then post-processing procedure is performed to distill the final secure keys.

\subsection{Airborne QKD performance evaluation}

\begin{figure}[h!]
    \centering
    \includegraphics[width=1\textwidth]{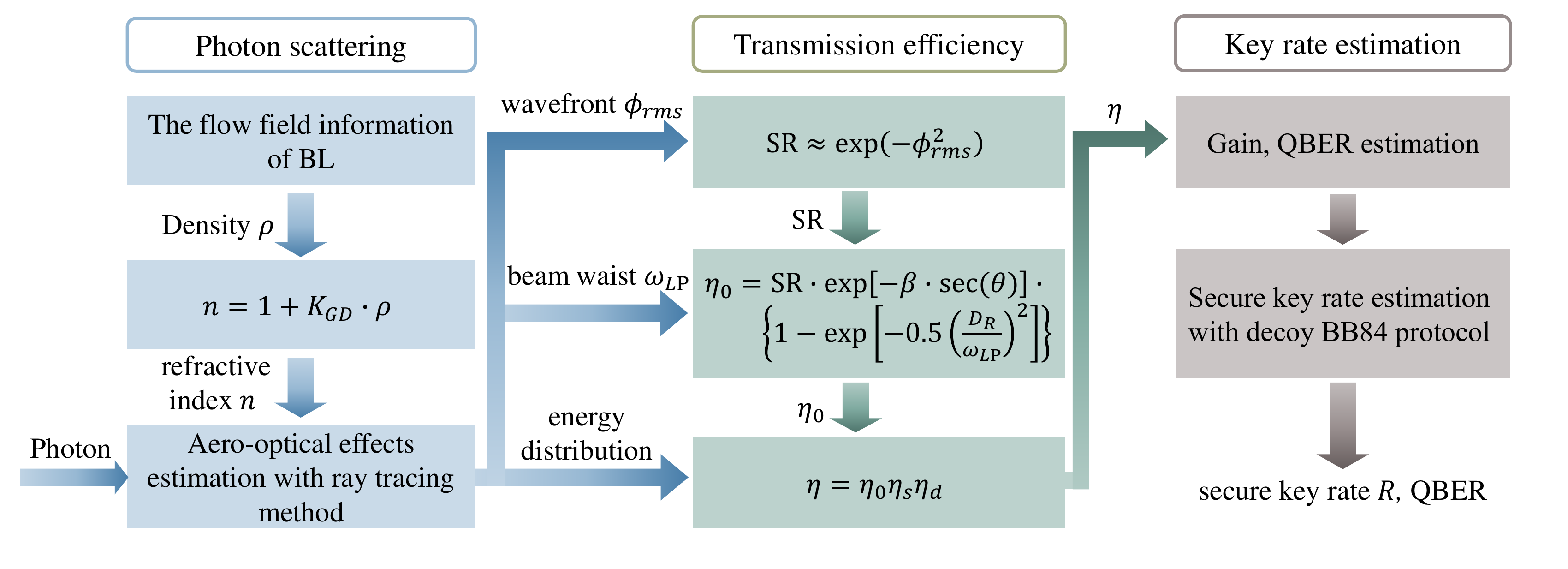}
    \caption{\label{fig3}Diagram of the airborne QKD performance evaluation procedure }
\end{figure}

The performance evaluation procedure for airborne QKD scheme is mainly contains three steps: photon scattering evaluation, transmission efficiency calculation and key rate estimation, shown in Figure~\ref{fig3}.

\subsubsection{Photon scattering.} Given the aircraft specification, speed $v$, relative flying height $h$ and the air density $\rho_h$, the density field distribution of the boundary layer can be simulated by the computational fluid dynamics software (such as CFX, Fluent, star-CD and comsol). Afterwards, the refractive index field can be obtained with Gladstone-Dale equation. Thus, when the Gaussian mode beam is propagating through the boundary layer to the ground station, aero-optical effects of photons can be evaluated with the ray tracing method. 

The normalized intensity of arrived photons at the ground station can be expressed as
\begin{equation}
   I(r, l)=\frac{2}{\pi \cdot \omega_{L P}} \exp \left(\frac{-2 r^{2}}{\omega_{L P}^{2}}\right),
\end{equation}
where $r$ is the radius of the beam, $l$ is the propagated distance of the photons.

$\omega_{LP}$ is the effective beam waist of the downlink photon at the ground station~\cite{RN118}
\begin{equation}
  \omega_{L P}=\sqrt{\omega_{L}^{2}+\left(\sigma_{T} \cdot l\right)^{2}},
\end{equation}
where $\sigma_{T}$ is the pointing error of the transmitter telescope. 

$\omega_{L}$ is the beam waist at the ground station prior to pointing errors
\begin{equation}
  \omega_{L}=l \frac{\lambda}{\pi \cdot \omega_{0}}\left[1+0.83 \cdot \sec (\theta)\left(\frac{D_{T}}{r_{0}}\right)^{5 / 3}\right]^{3 / 5},
  \end{equation}
where $r_{0}$ is the fried parameter in zenith~\cite{RN118}, $\theta$ is zenith angle of the receiving telescope and $D_T$ is the diameter of the transmitter telescope. $\omega_{0}$ is the waist radius of transmitted Gaussian beam
\begin{equation}
  \omega_{0}=0.316 D_{T},
  \end{equation}
where 0.316 results from the fact any aperture passed by real beam results in an Airy disk pattern~\cite{RN118}. 

Therefore, the aero-optical effects of transmitted photons can be calculated with the ray tracing method and equations~(\ref{equ_n})-(\ref{equ_phi}).

\subsubsection{Transmission efficiency.}

When the beam propagates through the boundary layer and illuminates the receiving telescope, the transmission efficiency $\eta_0$ can be calculated as~\cite{RN118}
\begin{equation}
  \eta_{0}=\mathrm{SR} \cdot \exp [-\beta \cdot \sec (\theta)] \cdot\left\{1-\exp \left[-0.5\left(\frac{D_{R}}{\omega_{L \mathrm{P}}}\right)^{2}\right]\right\},
\end{equation}
where $D_{R}$ is the diameter of the receiving telescope, and $\beta$ is the extinction optical thickness between sea level and altitude. The Strehl ratio ($\mathrm{SR}$) is the on-axis beam intensity at the target (far field receiver), $I_r$, divided by the intensity for a perfect on-axis intensity, $I_0$, at the target, with the $Mar\acute{e}chal$ approximation~\cite{RN124}
\begin{equation}
  \mathrm{SR} = \frac{I_r}{I_0} \approx \exp \left(-\phi_{\mathrm{rms}}^{2}\right)=\exp\left[-\left(\frac{2\pi OPD_{\mathrm{rms}}}{\lambda}\right)^2\right].
\end{equation}

With the perfect air condition and low flight speed (usually $|\vec{v}|\leq0.3$ Ma), airborne QKD will be performed without boundary layer effects, which results $\mathrm{SR}\approx 1.0$.

\subsubsection{Secure key rate estimation.} In the airborne QKD system, the photon transmission efficiency $\eta$ will be decreased, with the aero-optical effects of the aircraft boundary layer, which can be calculated as
\begin{equation}
    \eta = \eta_0 \eta_s \eta_d,
\end{equation}
where $\eta_s$ is the system receiving efficiency caused by constant optical components and $\eta_d$ is the detector efficiency.

Thus, the decreased overall gain $Q_\mu$ and the increased overall QBER $E_\mu$ can be calculated by equation (\ref{equ_Qmu}) and (\ref{equ_Emu}). Also, $Q_1$ and $e_1$ of single photon counts can be estimated by equation (\ref{equ_Q1}) and (\ref{equ_e1}). Afterwards, the secure key rate $R$ can be obtained by equation (\ref{equ_R}).

\begin{table}[h!]
	\caption{\label{tab_para}Parameters of airborne QKD.}
	\footnotesize
	\begin{tabular}{@{}llll}
		\hline
		\textbf{Payload} & \textbf{Parameter} & \textbf{Description} & \textbf{Value} \\
		\hline
		& $v$ & Flight speed & 0.7~Ma \\
		& $h$ & Relative flying height & 10~km \\
		& $\rho _h$ & Air density & 0.413~kg/m$^{3}$ \\
		\multirow{-4}{*}{\textbf{Aircraft}} & $d$ & \begin{tabular}[c]{@{}l@{}}The shortest horizon distance between\\ the aircraft and the ground station\end{tabular} & 10~km \\ \hline
		& $D_T$ & Diameter of the transmitter telescope & 0.05~m \\
		& $\delta _T$ & Transmitter pointing precision~\cite{RN85} & 150~\textmu rad \\
		& $\lambda$ & Transmitter wavelength & 1550~nm \\
		& $\omega_0$ & Waist radius & 0.0158~m \\
		\multirow{-5}{*}{\textbf{Photon Source}}
		
		& $r_0$ & Fried parameter in zenith~\cite{RN118} & 0.2~m \\ \hline
		& $D_R$ & Diameter of the receiver telescope & 0.3~m \\
		& $e_d$ & System detection error rate & 1\% \\
		& $p_d$ & Dark count rate & $2\times10^{-6}$ \\ 
		& $\eta_d$ & Detector efficiency & 15\% \\
		\multirow{-5}{*}{\textbf{Ground station}} & $\eta_{s}$ & Receiving optical module efficiency & 60\% \\ 
		\hline
		& $\mu$ & Intensity of signal states & 0.8 \\
		& $\nu$ & Intensity of decoy states & 0.1 \\
		& $N$ & System repetition rate & 100~MHz \\
		& $P_s$ & Probability of signal states & 50\% \\
		& $P_d$ & Probability of decoy states & 25\% \\
		\multirow{-6}{*}{{ \textbf{Protocols}}} & $P_v$ & Probability of vacuum states & 25\% \\ \hline
	\end{tabular}
\end{table}

\begin{figure}[h!]
      \centering
      \includegraphics[width=0.9\textwidth]{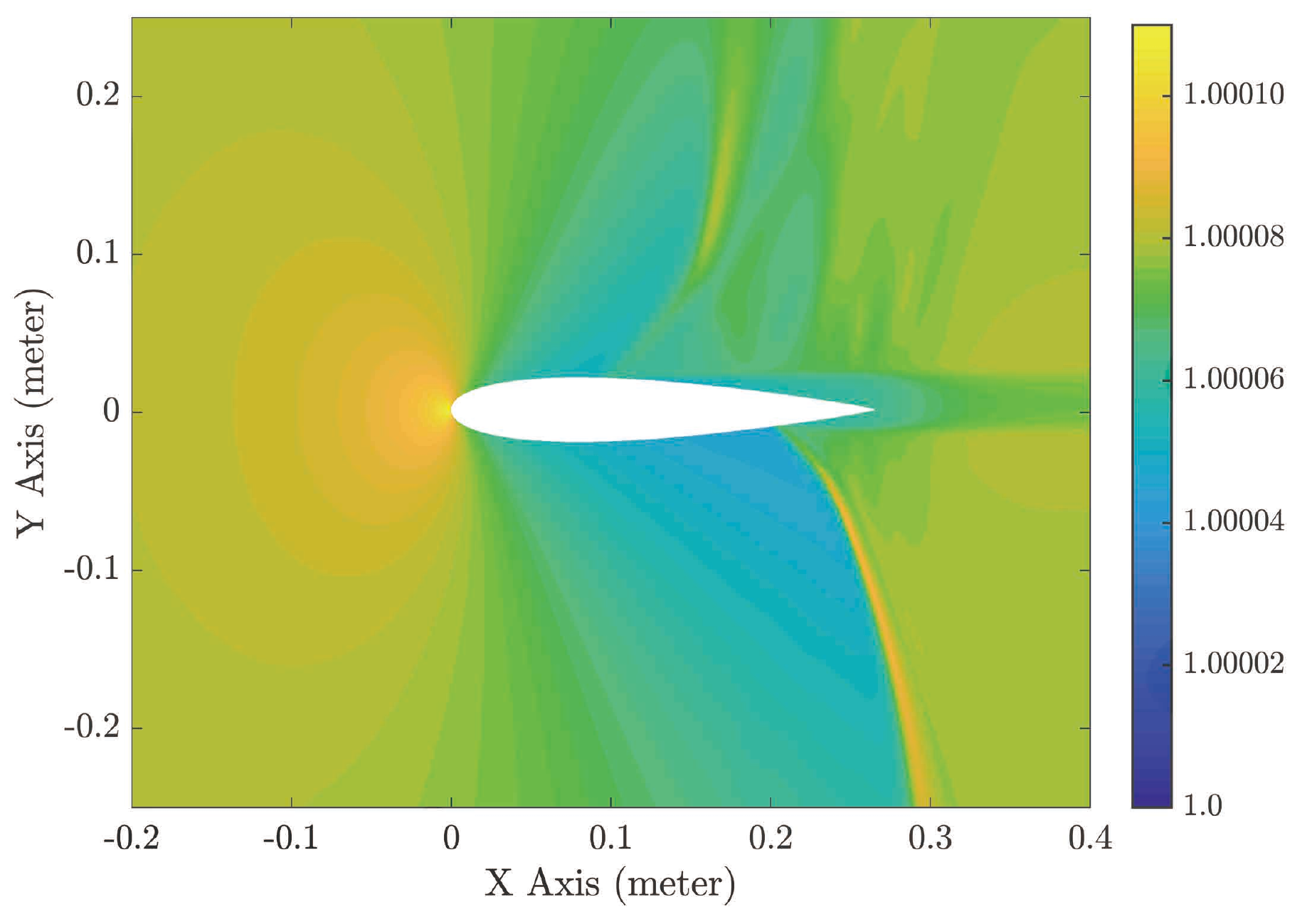}
      \caption{\label{fig4}The evaluated refractive index distribution of the NACA0015 airfoil boundary layer.}
\end{figure}

\section{Performance analysis}

In previous airborne QKD implementations, the quantum photon source payload is usually mounted in the belly pod of aircraft, where the airflow turbulence of boundary layer is much heavier than the airfoil. In this article, we mount the quantum photon source in the airfoil, as slight aero-optical effects to be tolerated and lots of standard models are specified.

The specific parameters of the aircraft, quantum photon source payload, and optical ground station are shown in Table~\ref{tab_para}. Here, the standard ``NACA0015'' airfoil is chosen for the performance analysis of our specified airborne QKD system.

Given the detailed aircraft description with $v=\SI{0.7}{Ma}$, the boundary layer will be generated around the NACA0015 airfoil and its density field distribution can be simulated by the computational fluid dynamics software (CFX). Afterwards, the refractive index distribution can be calculated by equation (\ref{equ_n}), shown in Figure~\ref{fig4}.

\begin{figure}[h!]
      \centering
      \includegraphics[width=0.5\textwidth]{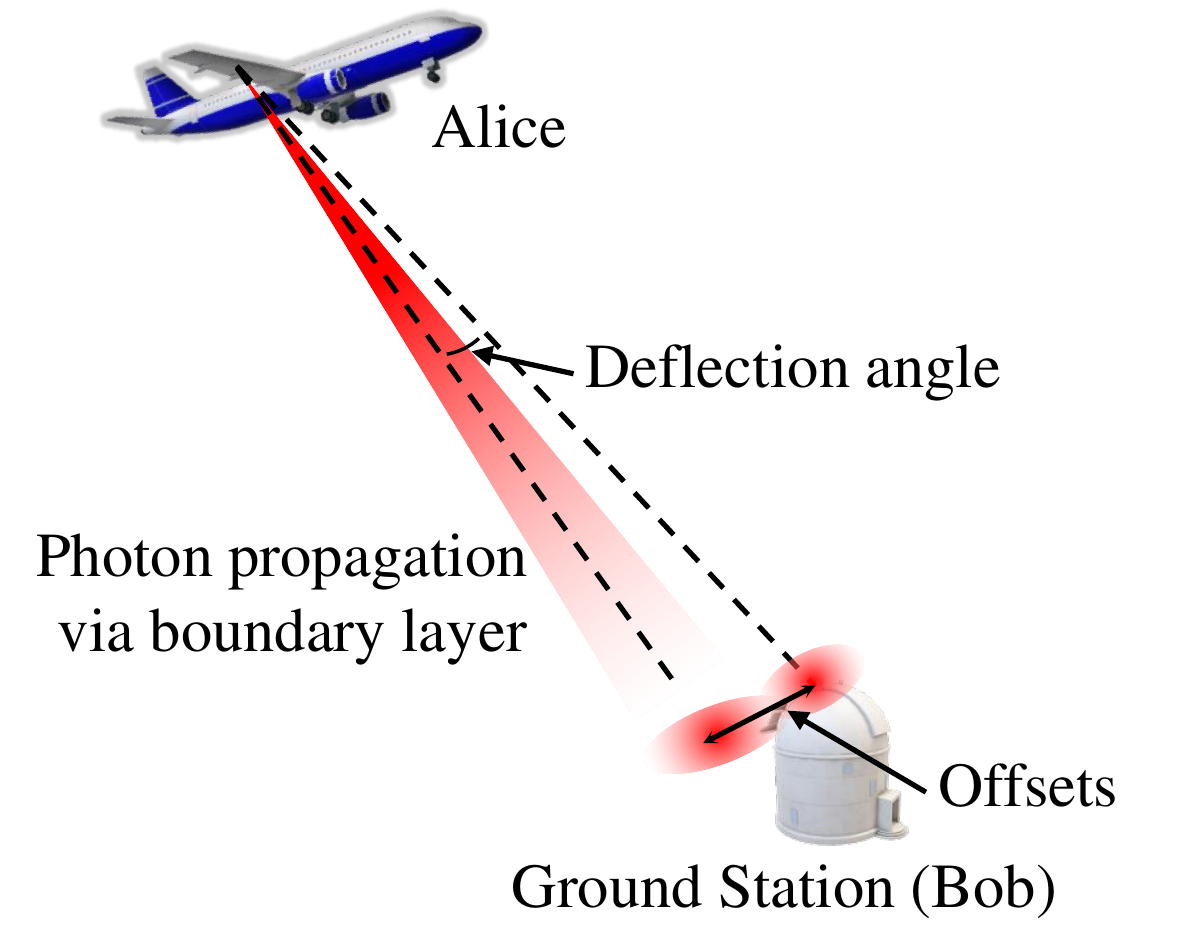}
      \caption{\label{fig5}Schematic diagram of photon propagation with a deflection angle via the boundary layer.}
\end{figure}

\begin{figure}[h!]
      \centering
      \includegraphics[width=0.9\textwidth]{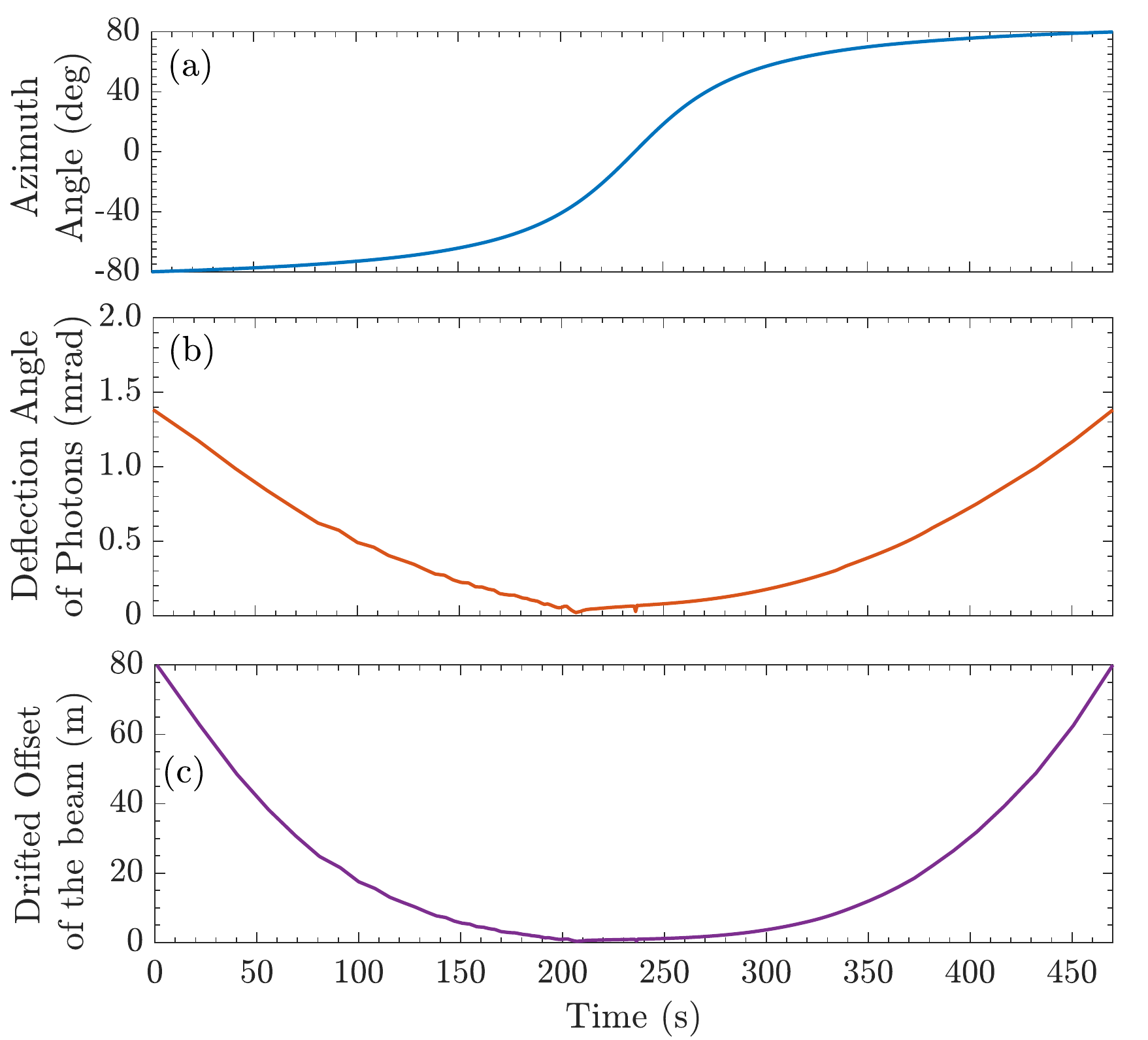}
      \caption{\label{fig6}(a) The azimuth angle $\alpha$ over the flight time. (b) The deflection angle of transmitted photons. (c) The drifted offset of the beam, which reaches to the ground station.}
\end{figure}

 In the air-to-ground QKD scenario, the photon propagated via the boundary layer would be deflected with a certain angle, as shown in Figure~\ref{fig5}. With the settings in Table~\ref{tab_para}, the evaluated deflection angle and the drifted offset of the beam, which reached to the ground station, are shown in Figure~\ref{fig6}.
 
The deflection angle caused by the boundary layer can be up to \SI{1.4}{mrad}, and the correspondingly drifted offset of the beam at the ground station is about \SI{80}{meters}.  Therefore, pre-compensation strategies of the deflection angle, have to be performed to the acquisition, tracking, and pointing (ATP) module of the aircraft quantum photon source payload, with the evaluation of the aero-optical effects caused by the boundary layer. Meanwhile, the ATP performance can be further improved with extra beacon laser module.

\begin{figure}[h!]

      \includegraphics[width=0.98\textwidth]{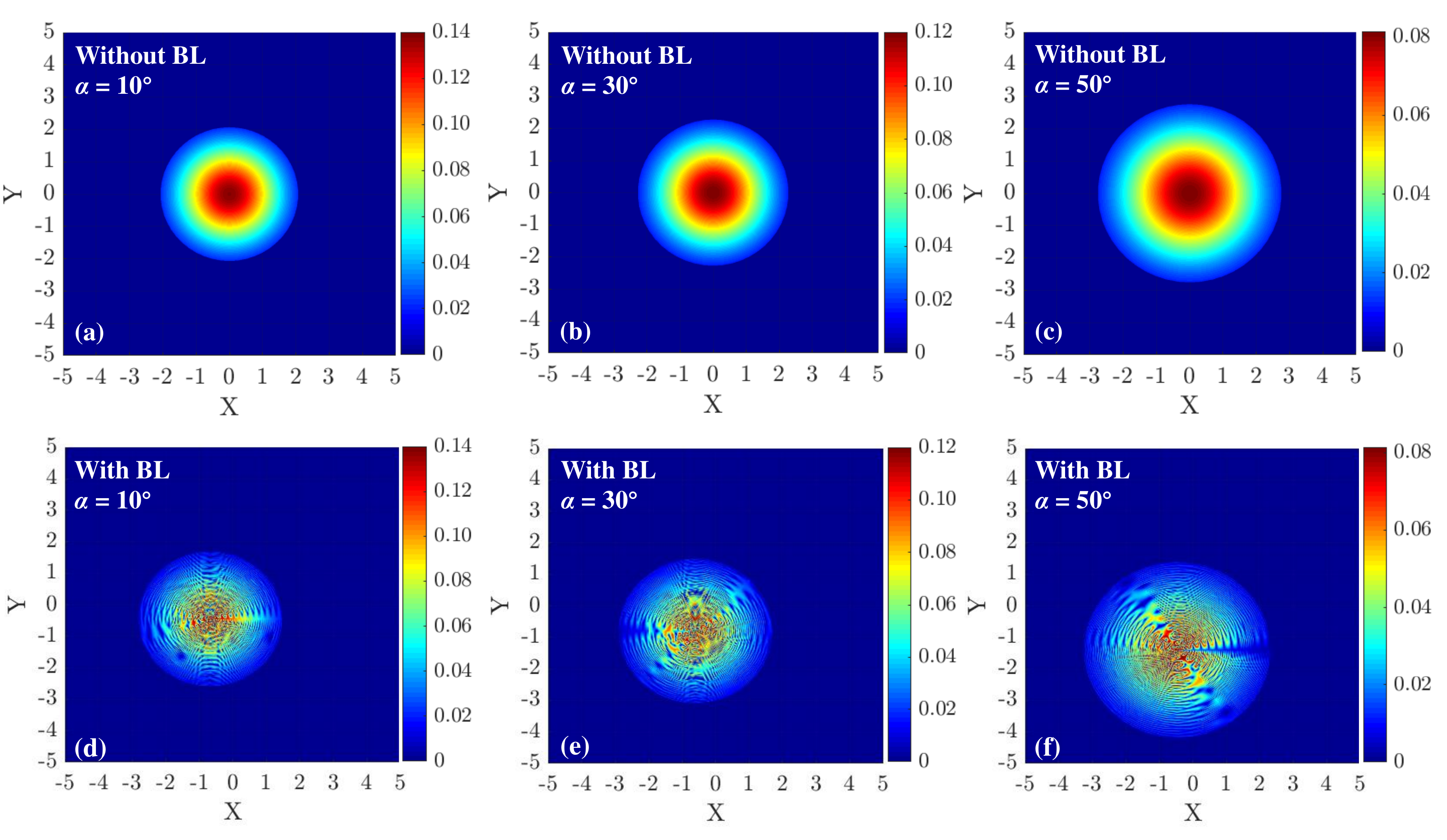}
      \caption{\label{fig7}(a)-(c) show the intensity distribution of the beam propagated without the boundary layer effects. (d)-(f) show the intensity distribution of the beam propagated through the boundary layer. Here $\alpha=10^\circ,30^\circ,50^\circ$, and the unit of axis is meters.}
\end{figure}

The random wavefront aberration of the photons, which are propagated through the boundary layer, is a huge challenge to the airborne QKD system. In Figure~\ref{fig7}, we show the energy distribution of photons received by the ground station with different azimuth angles by estimating the $OPL$ and $OPD$.

As shown in Figure~\ref{fig7}, the wavefront aberration and the diffusion of the beam are much heavier with larger azimuth angle. Meanwhile, the wavefront aberration caused by the boundary layer is complicated and harder to predict in the realistic airborne QKD scenario. Thus, the beacon laser module on the aircraft and the adaptive optics (AO) module on the ground station are suggested to be employed. Therefore, the wavefront aberration of quantum signals can be compensated by adjusting the AO module, based on the analyzed results of the beacon beam wavefront.

\begin{figure}[h!]
      \centering
      \includegraphics[width=0.9\textwidth]{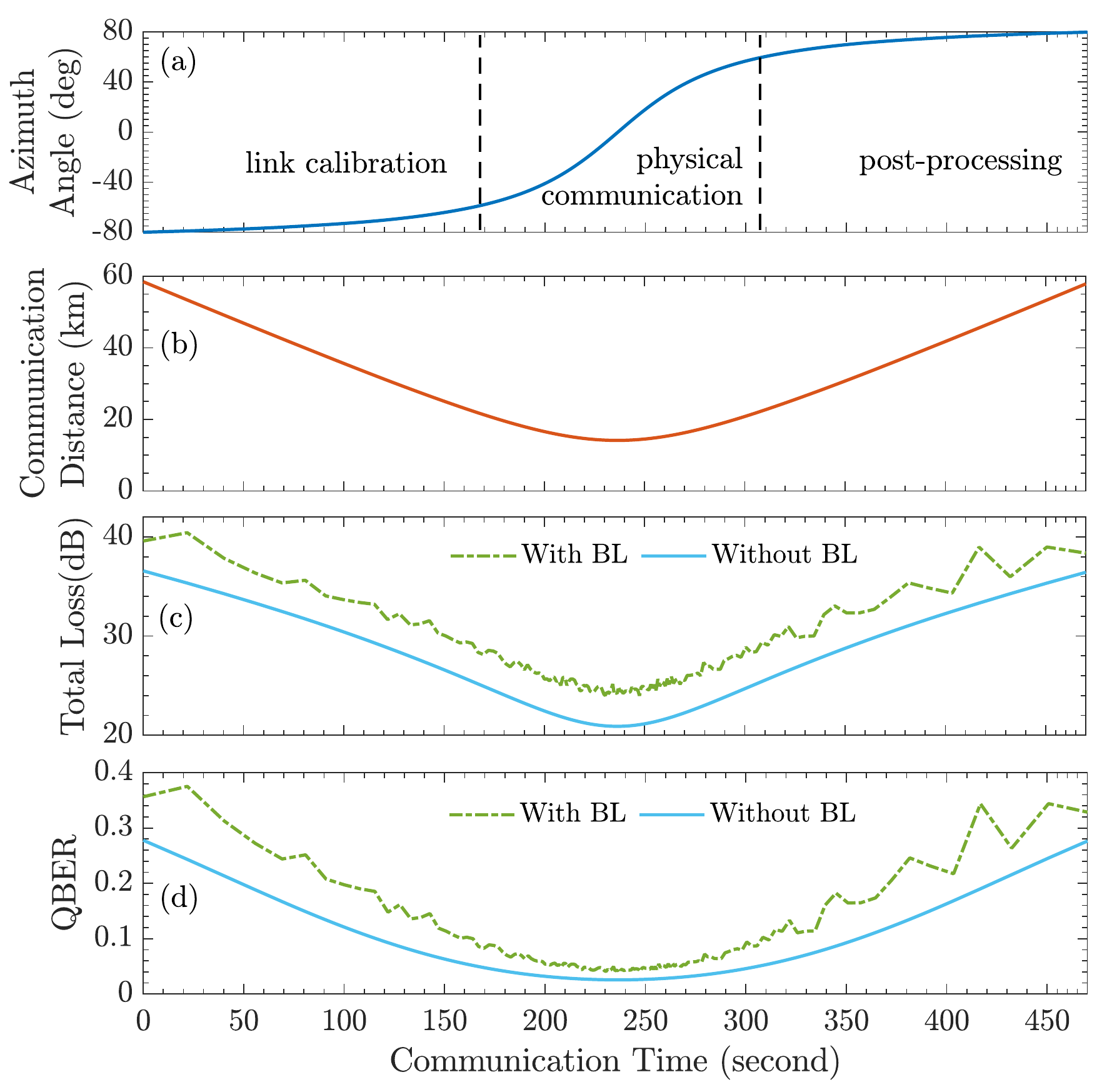}
      \caption{\label{fig8}(a) Azimuth angle over the flight time. (b)The communication distance between the aircraft and the ground station. (c) The total loss in the airborne QKD scenario. (d) The estimated QBER over the flight time.}
\end{figure}

Finally, the performance of the whole airborne QKD session is evaluated and the result is shown in Figure~\ref{fig8}. The total communication time is around \SI{470}{seconds} and the communication distance between the aircraft and the ground station is around \SI{15}{km} to \SI{60}{km}. The boundary layer around aircraft will introduce around \SI{3.5}{dB} channel loss to the transmitted photons, as shown in Figure~\ref{fig8}(c). Once the azimuth angle $|\alpha|\geq 60 ^\circ$, the estimated QBER of signal states will be larger than \SI{10}{\%}, which would result in no secure keys, as shown in Figure~\ref{fig8}(d). Thus, we perform the link calibration procedure with azimuth angle $\alpha \leq -60 ^\circ$ and the post-processing procedure with $\alpha \geq 60 ^\circ$, shown in Figure~\ref{fig8}(a). Therefore, the total quantum communication time is around \SI{140}{seconds} and the communication distance is around \SI{20}{km} to \SI{30}{km}, the estimated final secure key rate is around \SI{386.4}{bps}. If there's no boundary layer surrounds the aircraft, the estimated secure key rate would be around~\SI{1.32}{kbps}. In summary, the boundary layer effects can not be ignored in the airborne QKD scenario and heavily decreases the final secure key rate.

\section{Conclusion} 

Airborne quantum key distribution (QKD) will be a flexible bond between terrestrial fiber QKD network and the quantum satellites, which can establish a mobile, on-demand and real-time coverage quantum network. However, the randomly distributed boundary layer is always surrounded to the surface of the aircraft, which would introduce random wavefront aberration, jitter and extra intensity attenuation to the transmitted photons between the aircraft and the ground station. In this article, we propose the detailed performance evaluation scheme of airborne QKD with boundary layer effects. The analyzed photon deflection and wavefront aberration results show that the aero-optical effects caused by the boundary layer can not be ignored, which would heavily decrease the final secure key rate. In our proposed airborne QKD scenario, the boundary layer would introduce $\sim$\SI{3.5}{dB} loss to the transmitted photons and decrease $\sim$\SI{70.7}{\%} of the secure key rate. With tolerated quantum bit error rate set to \SI{10}{\%}, the suggested quantum communication azimuth angle between the aircraft and the ground station is within $60^\circ$. Furthermore, the optimal beacon laser module and adaptive optics module are suggested to be employed to improve the performance of airborne QKD system. Our detailed airborne QKD evaluation study can be performed to the future airborne quantum communication designs.

\section*{Appendix}

\begin{backmatter}

\section*{Funding}
This work is supported by the National Natural Science Foundation of China under Grant No.61971436, No.61972410 and No.61803382, the Natural Science Basic Research Plan in Shaanxi Province of China (No.2018020JQ6020), the Research Plan of National University of Defense Technology under Grant No.ZK19-13 and No.19-QNCXJ-107, the Postgraduate Scientific Research Innovation Project of Hunan Province under Grant (No.CX20200003).

\section*{Availability of data and materials}
Simulations and scripts are made available upon request by the corresponding author BL.

\section*{Competing interests}
The authors declare that they have no competing interests.

\section*{Authors' contributions}
The model design and simulations were done by HCY and BYT and BL. The background was analysed by HC and YX. JT and WRY handled the verification of manuscript. The effort was conceived and supervised by BL and co-supervised by LS. HCY and BL wrote the draft and all authors reviewed the manuscript..

\bibliographystyle{manuscript} 
\bibliography{ref}     
\end{backmatter}
\end{document}